\documentclass[aps,prd,preprint,groupedaddress,showpacs,A4paper]{revtex4-1}

\usepackage{amsmath,amssymb}
\usepackage{hyperref}
\usepackage{tikz}
\usetikzlibrary{intersections}
\usetikzlibrary{decorations.pathmorphing}
\usetikzlibrary{decorations.markings}

\tikzset{
    laser/.style={decorate, decoration={snake,segment length=5mm}, draw=black},
    electron/.style={draw=black, postaction={decorate},
        decoration={markings,mark=at position .55 with {\arrow[draw=black]{>}}}},
}

\newcommand{\beq}{\begin{equation}}
\newcommand{\eeq}{\end{equation}}
\newcommand{\bea}{\begin{eqnarray}}
\newcommand{\eea}{\end{eqnarray}}

\newcommand{\phiV}{\phi_{_{\mathrm{V}}}}
\newcommand{\aV}{a_{_{\mathrm{V}}}\!}

\newcommand{\bdV}{b^\dag_{_{\mathrm{V}}}\!}

\newcommand{\kx}{k{\cdot}x}

\newcommand{\intph}{\int\!\frac{{\bar{}\kern-0.45em d}^{\,3}{p}}{2E_{p}}}
\newcommand{\intp}{\int\!\frac{{\bar{}\kern-0.45em d}^{\,3}{p}}{2E^*_p}}
\newcommand{\intq}{\int\!\frac{{\bar{}\kern-0.45em d}^{\,3}{q}}{2E^*_q}}

\newcommand{\ee}{\mathrm{e}}

\newcommand{\bra}[1]{\langle #1|}
\newcommand{\ket}[1]{|#1\rangle}
\newcommand{\braV}[1]{{_{_{\mathrm{V}}}}\!\bra{#1}}
\newcommand{\ketV}[1]{\ket{#1}_{_{\mathrm{V}}}}

\newcommand{\ccdot}{{\cdot}}
\begin{document}


\title{Sideband Mixing in Intense Laser Backgrounds}


\author{Martin~Lavelle and David~McMullan}
\affiliation{School of Computing and Mathematics\\ University of Plymouth \\
Plymouth, PL4 8AA, UK}

\date{\today}

\begin{abstract}
The electron propagator in a laser background has been shown to be made up of a series of sideband poles. In this paper we study this decomposition by analysing the impact of the residual gauge freedom in the Volkov solution on the sidebands. We show that the gauge transformations do not alter the location of the poles. The identification of the propagator from the two-point function is maintained but we show that the sideband structures mix under residual gauge transformations.
\end{abstract}



\maketitle

\section{Introduction}
The recent rapid progress in laser technologies  offers a timely testing ground for quantum field theory techniques associated with non-trivial backgrounds~\cite{DiPiazza:2011tq}. In this paper we are going to study charge propagation in such a background. A novel feature of a propagating charge in a laser  is that it is indistinguishable from a charge which absorbs a given number of laser photons and also emits the same number of photons degenerate with the laser. Such laser induced degeneracies  have a close parallel with the soft and collinear degeneracies associated with the infra-red regime in both QED and QCD~\cite{Bloch:1937pw}\cite{Kinoshita:1962ur}\cite{Lee:1964is}\cite{Lavelle:2005bt} while the induced mass effects in a laser background should help to refine our understanding of the current versus constituent mass distinction in QCD~\cite{Lavelle:1995ty}. Needless to say, the new generation of laser facilities that will soon be on line will help to both inform and refine our  theoretical  understanding of these vexed issues.

In QED we usually expand around the free theory but in a laser background we use the interacting, Volkov solution~\cite{Volkov:1935zz}\cite{Heinzl:2011ur} as our starting point. This solution is much richer than in the normal perturbative vacuum and, as we will summarise below, alters the propagator which becomes a sum of so-called sideband poles~\cite{Reiss:1962}\cite{Reiss:2009ed}. As this is not a free theory, we have to address the effects of local gauge transformations~\cite{Heinzl:2008rh} on this solution and the propagator.

We recall~\cite{Volkov:1935zz} that the  solutions of a scalar field in a plane wave background are distorted. For a linearly polarised background
where the vector potential is given by
\begin{equation}\label{pot}
A_\mu(x)=a_\mu\cos (\kx)\,,
\end{equation}
and where the constant amplitude $a_\mu$ is space-like and taking the null vector $k^\mu$ to be spatially aligned along the laser direction, the matter field is described by
\begin{equation}\label{phiV}
\phiV(x)=\intp \left(  D(x,p)\aV(p)+D(x,-p)\bdV(p)\right)\,,
\end{equation}
where
\begin{equation}
D(x,p)=\ee^{-ip\cdot x}\ee^{i\left(eu\sin(\kx)+e^2v\sin(2\kx)\right)}\,,
\end{equation}
and
\begin{equation}\label{potbits}
u=-\frac{p\ccdot a}{p\ccdot k}\qquad \mathrm{and}\qquad v=\dfrac{a^2}{8p\cdot k}\,.
\end{equation}
In this expression the momentum $p$ is on-shell at $m_{\star}$ where the laser shifted mass~\cite{Reiss:1962}\cite{Brown:1964zz}\cite{Nikishov:1964zza} \cite{Harvey:2012ie}\cite{Kohlfurst:2013ura} is
\begin{equation}
p^2={m_{\star}}^2=m^2-\tfrac12 e^2 a^2\,.
\end{equation}
In this paper we do not explicitly distinguish  between on-shell and off-shell momenta as it has no impact on our discussion of gauge dependence. See~\cite{Lavelle:2013wx} for a fuller discussion.

We recall further that (\ref{phiV})  may be written as a sum over modes
\begin{equation}\label{summingmodes}
\phiV(x)=\sum_n \phi_n(x)\,,
\end{equation}
where
\begin{equation}\label{174174}
 \phi_n(x)= \intp \left( \ee^{-ie p\ccdot x}\ee^{in\kx} J_n(eu,e^2v)\aV(p)+
\ee^{ie p\ccdot x}\ee^{in\kx} J_n(eu,-e^2v)
\bdV(p)\right)\,,
\end{equation}
and the generalised Bessel function, $J_n(eu,e^2v)$, is defined in terms of Bessel functions via
\begin{equation}\label{genBsfndef}
J_n(eu,e^2v)=\sum_r J_{n-2r}(eu)J_r(e^2v)
\,.
\end{equation}

The Volkov propagator contains not just the standard pole $i/(p^2-{m_*}^2)$ familiar from perturbation theory but also infinitely many sideband poles
of the form $i/((p+nk)^2-{m_*}^2)$ where $n$ is any integer~\cite{Brown:1964zz}\cite{Eberly:1966b}\cite{Reiss:1966A}\cite{Ehlotzky:1967c}\cite{Ilderton:2012qe}.
As we have previously seen~\cite{Lavelle:2013wx}, the propagator is not given by the two-point function of the full Volkov field but is identified as the diagonal part of the two-point function in the vacuum $\ketV 0$ picked out by the Volkov annihilation operators:
\begin{equation}\label{propdiagdef}
i D_V(x-y)=\sum_n\braV 0 T \phi_n(x)\phi^{\dag}_n(y)\ketV 0\,.
\end{equation}
This is to ensure that the propagator represents processes where there is a fixed momentum flow through the matter field. This can also be understood~\cite{Lavelle:2013wx} in terms of degenerate processes extending the Lee-Nauenberg~\cite{Lee:1964is}  characterisation of the infra-red problem~\cite{Lavelle:2005bt}.

The form of the vector potential chosen here requires that $k\cdot a=0$ which corresponds from~(\ref{pot}) to a Landau like gauge as $\partial_\mu A^\mu=0$. In~\cite{Lavelle:2013wx} the propagator was constructed in this gauge. The mass shift and wave function renormalisation were  calculated to all orders in an operator formalism. This was further verified to the first few orders by explicit diagrammatic calculations. Each term in the sum~(\ref{propdiagdef}) generates a separate, so-called sideband structure:
\begin{equation}\label{propdiagdef2}
\int d^4x \, \ee^{-ip\cdot (x-y)}\braV 0 T (\phi_n(x)\phi^{\dag}_n(y))\ketV 0 = \frac{Z_2^{(n)}(u,v)}{(p+nk)^2-{m_*}^2+i\epsilon}\,,
\end{equation}
showing the common mass shift and distinct wave function renormalisations of the sidebands. Here $p$ is off-shell.
It has been  argued that the central sideband, corresponding to $n=0$ and produced by the $\phi_0(x)$ mode, may dominate in some regimes~\cite{Eberly:1966b}.
In this paper we want to address the issue of the residual gauge freedom which is opened up by the boundary conditions imposed on the plane wave laser background.

Below we will show that, although the Volkov field transforms with the expected phase shift characteristic of a charged matter field under such a residual gauge transformation, the modes (\ref{summingmodes}) actually mix with each other in a non-trivial manner. This mixing of the modes raises a question about whether the above identification of the propagator~(\ref{propdiagdef}) is consistent with gauge transformations. We will demonstrate below that the construction of the propagator is robust under such transformations and that the overall effect of gauge transformations may be absorbed into shifts of the wave function renormalisation factors.

We therefore now turn to the gauge freedom in the Volkov formalism, its effects on the various modes of the Volkov field and thus build up the diagonal sum~(\ref{propdiagdef}). Although our conclusions hold to all orders, we shall, for illustrative purposes, demonstrate them perturbatively.

\section{Residual Gauge Transformations}
There is in the gauge fixing discussed above a residual gauge freedom as we can make the replacement
\begin{equation}
A_\mu(x)\to A_\mu(x)+\partial_\mu \lambda(x)\,,
\end{equation}
where $\lambda(x)=\lambda\sin( \kx)$ and $\lambda$ is a constant. This corresponds to the amplitude shift
\begin{equation}
a_\mu\to a_\mu+\lambda k_\mu\,,
\end{equation}
which still preserves our gauge choice due to the null nature of $k^\mu$.
Under this transformation we have, from (\ref{potbits}),
\begin{equation}\label{potbitstrans}
u\to u-\lambda\qquad \mathrm{and}\qquad v\to v\,.
\end{equation}%
Similarly the distortion factor transforms as
 \begin{equation}
D(x,p)\to \ee^{-ie\lambda(x)}D(x,p)\,.
\end{equation}
From (\ref{phiV}) we see the phase shift
 \begin{equation}\label{131474}
\phiV(x)\to \ee^{-ie\lambda(x)} \phiV(x)\,,
\end{equation}
as would be expected of a charged matter field under gauge transformations. This is a local gauge transformation and, as the field extends to spatial infinity along the laser direction, the transformation does not vanish asymptotically along the laser. This residual gauge transformation is consistent both with our original Landau gauge condition and the boundary conditions of the Volkov solution.

We now want to analyse the impact of the gauge freedom on the various modes of the Volkov field. As the propagator is constructed from the diagonal sum over the modes (\ref{propdiagdef}) it is crucial that we know how they transform. In (\ref{174174}) the generalised Bessel functions, through their dependence on $u$, are responsible for the gauge dependence of the fields
\begin{equation}\label{Jchange}
J_n(eu,e^2v)\to J_n(e(u-\lambda),e^2v)=
\sum_m J_m(e\lambda) J_{n+m}(eu,e^2v)
\,,
\end{equation}
where we used (\ref{potbitstrans}). This means that the Volkov modes mix under such a local gauge transformation as
\begin{equation}\label{phinchange}
\phi_n(x)\to
\sum_s J_s(e\lambda) \phi_{n+s}(x) \ee^{-is\kx}
\,,
\end{equation}
with a Bessel function dependent weighting. It is useful here to verify that this is consistent with the overall transformation of the Volkov field. From (\ref{summingmodes}) we have
\begin{equation}\
\phiV(x)\to
\sum_n \sum_s J_s(e\lambda) \phi_{n+s}(x) \ee^{-is\kx}
\,.
\end{equation}
Shifting the label $n$ and using the standard result
\begin{equation}\
\ee^{i\ell\sin(\kx)}=\sum_r\ee^{ir\kx}J_r(\ell)\,,
\end{equation}
we find that the Volkov field
\begin{equation}
\phiV(x)\to
\ee^{-ie\lambda(x)} \sum_r\phi_{r}(x)
\,,
\end{equation}
which shows that, as expected from (\ref{131474}), the phase may be extracted from the sum over modes.

It is clear from (\ref{phinchange}) that there is a mixing of the modes and it is not obvious that the identification of the propagator as a diagonal sum is compatible with this mixing. To study this perturbatively, let us consider the lowest modes in the propagator. It will be apparent how this extends to higher orders. The first few terms in the diagonal sum around the central term ($n=0$) are:
\begin{equation}
\braV 0 T (\phi_0(x)\phi_0^{\dagger}(y)+\phi_1(x)\phi_1^{\dagger}(y)+\phi_{-1}(x)\phi_{-1}^{\dagger}(y)+\dots\ketV 0
\,.
\end{equation}
Now from (\ref{phinchange}) and the standard series representation of the Bessel function
we see that at order $e^2$ the only term that is gauge dependent is $\phi_0(x)\phi_0^{\dagger}(y)$. We find that
\begin{equation}\label{173974}
\phi_0(x)\to
\left(1-\tfrac14{e^2\lambda^2} \right)\phi_0(x)+\tfrac{e\lambda}2\phi_1(x)\ee^{-ik\ccdot x}-\frac{e\lambda}2\phi_{-1}(x)\ee^{ik\ccdot x}
\,,
\end{equation}
and similar for $\phi_0^{{\dagger}}$.

We see that the modes mix under a gauge transformation and that, more generally, expanding up to order $e^{2n}$ mixes modes whose labels are separated by $n$.  Furthermore, the factors of $\ee^{\pm ik\ccdot x}$ here might initially appear concerning as such a $k$-dependence in the diagonal two-point function would be incompatible with its interpretation as a propagator. However, we will see below that although the mixing is real this $k$-dependence cancels in our result and the propagator interpretation of the diagonal sum~(\ref{propdiagdef}) holds.

We now want to express the modes in terms of the Volkov creation and annihilation operators. Rewriting the generalised Bessel functions in (\ref{174174}) in terms of Bessel functions via (\ref{genBsfndef}) yields
\begin{equation}
J_0(eu,e^2v)= J_0(eu)J_0(e^2v)+J_2(eu)J_{-1}(e^2v)+J_{-2}(eu)J_{1}(e^2v)+\dots
\,.
\end{equation}
The second and third terms on the right hand side here are of order $e^4$, so to order $e^2$
\begin{equation}
J_0(eu,e^2v)= 1-\frac{e^2}4u^2
\,.
\end{equation}
We thus obtain to leading order in the coupling
\begin{equation}\label{formphi0}
\phi_0(x)= \intp\left(
\ee^{-ip\ccdot x}\left(1-\frac{e^2}4u^2\right)a_V(p)
+
\ee^{ip\ccdot x}\left(1-\frac{e^2}4 u^2\right)b^{{\dagger}}_V(p)
\right)
\,.
\end{equation}
A similar calculation leads to
\begin{equation}\label{formofphipm1}
\phi_{\pm1}(x)= \mp\frac e2\intp\left(
\ee^{-ip\ccdot x}\ee^{\pm ik\ccdot x}\left(1-\frac{e^2}4u^2\right)a_V(p)
+
\ee^{ip\ccdot x}\ee^{\mp ik\ccdot x}\left(1-\frac{e^2}4u^2\right)b^{{\dagger}}_V(p)
\right)
\,.
\end{equation}
It is important to notice how the factors of $\ee^{\pm ik\ccdot x}$ enter here. This means that such factors cancel on substitution into~(\ref{173974}). We find from (\ref{173974}), (\ref{formphi0}) and (\ref{formofphipm1}) that  the mode $\phi_0(x)$, expressed in terms of creation and annihilation operators, up to order $e^2$ transforms under the residual gauge transformation as
\begin{equation}\label{1739742}
\phi_0(x)\to \intp
\left(
1-\frac{e^2}4(u-\lambda)^2
\right)\left(\ee^{-i p\ccdot x} a_V(p)+\ee^{ip\ccdot x}b^{\dag}_V(p)\right)
\,.
\end{equation}
This last equation shows the leading order, local gauge transformations of the mode $\phi_0$. However, it should be emphasised that this result includes terms from  mixing with the modes $\phi_{\pm1}$ and will receive further contributions from $\phi_{\pm2}$ at order $e^4$ etc. It is clear that at all orders, all of the modes will mix under a gauge transformation. This demonstrates that restricting to specific modes is a gauge dependent, and unphysical approximation. The $e^{i\kx}$ factors have cancelled in (\ref{1739742}) and so if under our gauge transformation $\phi_n\to\tilde{\phi}_n$ the diagonal term $\braV 0 T (\tilde{\phi}_n(x)\tilde{\phi}^{\dag}_n(y))\ketV 0 $
continues to generate solely the sideband pole in $(p+nk)^2-{m^*}^2$, although $\tilde{\phi}_n$ itself contains contributions from all the other modes in the original gauge.

We have thus seen that under gauge transformations, propagators defined through the diagonal sum over modes remain as propagators. The changes affect the wave function renormalisation which acquires a $\lambda$-dependence. For the central sideband
\begin{equation}\label{132484}
Z_2^{(0)}(eu,e^2v)\to
1-\frac{e^2}2 (u-\lambda)^2+{\cal{O}}(e^4)\,.
\end{equation}
If we recall~\cite{Lavelle:2013wx} that the all orders wave function renormalisation constant for the $n$-th sideband in a linear background has the form
\begin{equation}\label{130084}
Z_2^{(n)}(eu,e^2v)=J^2_{n}(eu,e^2v)
\,,
\end{equation}
then~(\ref{132484}) corresponds to the shift
\begin{equation}\label{132684}
Z_2^{(0)}(eu,e^2v)\to Z_2^{0}(e(u-\lambda),e^2v)
\,.
\end{equation}
This exemplifies the gauge dependence of the wave function renormalisation and we reiterate that it is in part caused by a mixing of modes which previously generated other sideband states in the initial gauge. Note that although $Z_2^{(\pm1)}$ is gauge invariant at this order, it  becomes gauge dependent at the next order, and so on. The mass shift, as a potentially measurable quantity~\cite{Harvey:2012ie}, is gauge invariant as under our residual gauge tranformation
\[
a^2\to (a_\mu+\lambda k_\mu)(a^\mu+\lambda k^\mu)=a^2
\,,
\]
where we have used that $k^2=0$ and the Landau gauge $k\cdot a=0$. This is reflected in the expression for the mass shift which is only generated by diagrams with the four-point vertex and not by the gauge dependent three-point vertex in scalar QED. This is also the case for circular polarised backgrounds~\cite{Lavelle:2013wx}.  It should be contrasted with fermionic QED where the gauge dependent vertex is proportional to $e\slash \!\!\!a$ and it is possible to generate the gauge invariant structure $a^2$ via $\slash \!\!\!a \slash \!\!\!a$ from the fermionic vertex.

We have seen in this paper that the mass shift is gauge invariant and that the wave function renormalisation factors are gauge dependent. Furthermore, the different sidebands mix with each other under gauge transformations although, in the resulting gauge, each mode in the diagonal sum generates its corresponding sideband pole. This shows that in this theory we have a consistent identification of the propagator with its sideband structure.

The next step in understanding the quantum field theory of the Volkov solution is to identify the vertex structures in scalar QED in laser backgrounds. We expect similar structures in fermionic QED, however, this does introduce an additional complexity to the theory~\cite{Reiss:1966A} which needs to be revisited. Progress in these areas is essential in the development of the theory of scattering in laser backgrounds.

\bigskip

\noindent \textbf{Acknowledgements:} We thank Tom Heinzl and Ben King for helpful discussions.

\newpage



%

\end{document}